\documentclass[11pt]{article}
\usepackage{amsfonts}
\usepackage{graphicx}

\begin{document}

\title{The slippage paradox}
\author{Steffen BohnÊ\\
LPMA, UniversitŽ Paris Diderot (Paris 7) \& CNRS \\
Site Chevaleret, Case 7012\\
 75205 Paris Cedex 13, France}
\date{\today}

\maketitle
\begin{abstract}
Buying or selling assets leads to transaction costs for the investor. On one hand, it is well know to all market practionaires that the transaction costs are positive on average and present therefore systematic loss. On the other hand, for every trade, there is a buy side and a sell side, the total amount of asset and the total amount of cash is conserved. I show, that the apparently paradoxical observation of systematic loss of all participants is intrinsic to the trading process since it corresponds to a correlation of outstanding orders and price changes. 
\end{abstract}

\section{Introduction}
Market practionaires and academics are well aware of the importance of transaction costs when considering an investment. While for retail investors fees and commissions might be the determining factors, bigger financial players worry about what practionaires call slippage. Slippage is the difference between the asset price when the trading decision is taken, and the price actually realized 
by  a broker or an algorithmic execution system. \\
Let $\tilde  P$ the market price (last trade) when the (parent) order is passed to a broker or an algorithmic execution system. The order will then be executed in one or more trades. Let  $\Delta x_i$ be the exchanged quantity in the $i$th trade, and $P_i$ the corresponding price. Following the definition in \cite{Engle2006}, we shall define the slippage as
\begin{equation}
\label{eqn:defTR}
TR =  \sum_{\mathrm{trades} \, i} \Delta x_i (P_i -  \tilde  P)
\end{equation}
For a buy order, the slippage is the dollar amount of what the buyer pays in excess to an initial mark-to-market valuation. For a sell order, one sets $\Delta x_i < 0$ so that the slippages represents what the seller receives less that she would have (naively) expected.  In both cases, positive slippage presents therefore a loss to the investor. \\

Market practitioners know that slippage is positive on average, and that the loss associated with the slippage can easily wipe out the expected gain of an investment strategy when trading too much. Big institutions such as Morgan Stanley \cite{Engle2006} or Citigroup  \cite{Almgren2005} do dispose of private datasets with a sufficient large number of executed orders to derive models to predict the slippage as function of order size, volatility and other variables. Extensive academic and private
research focus of execution strategies thus to minimize the execution cost \cite{Kharroubi2009,Gatheral2010b, Almgren2001, Bertsimas1998}
\\ 
In spite of the empirical and theoretical knowledge about the transaction costs from the perspective of an individual agent, average positive slippage seems self-contradictory when we consider the market as a whole. Let there be a buyer $b$ and a seller $s$ starting to trade exclusively with each other. For each individual trade, $\Delta x _i^b = -\Delta x_i^s$. The trade prices $P_i$ are probably different from the common reference price $\tilde  P$, but they are the same for buyer and seller (no commissions of fees). From the definition above follows that $TR ^b + TR^s = 0$, the sum and thus the average transaction cost is strictly zero.  \\
Of course, In a more realistic setting,  there is more than one seller and more than one buyer, and their reference prices are not identical. Hence, one would naively expect the market average of the transaction costs not to be exactly zero, but converge to zero in the long run. I therefore consider the fact that the transaction cost are in general positive as a paradox.\\

One might tempted to identify the positive slippage of investors with then gains of market makers \cite{Kyle1985,Glosten1985} or high frequency traders. However, considering the magnitude of the cumulated slippages, these agents must then have unrealistic high earnings. As I will argue in section \ref{sec:slippage}, the presence of market makers is not necessary to explain the positive transaction costs. \\

Most of current research identify the transaction cost with the impact of trading. Buying an asset pushes the price upwards, while selling exerts a negative price pressure. The flaw of this familiar reasoning is that, for every trade, there is a buyer and a seller. If the buyer causes a price increase, the seller a price decrease, which is then direction of the price move? \\
Some empirical studies based on trade and tick data \cite{Lillo2003, Lillo2004,Bouchaud2004a, Bouchaud2004b} consider the impact of market orders. Stochastic orderbook models illustrate the mechanical effect of market orders \cite{Daniels2003}. Does this imply that execution based only on passive limit orders lead to negative slippage? \cite{Eisler2010} study price changes induced by placing and canceling limit orders. However, this microscopic approach is to complex to derive an execution cost of a limit order based execution strategies. In general, a rational execution algorithm alternates market and limit orders \cite{Rosu2009} thus to minimize the execution. To my knowledge, there exists no strategy for which the impact of execution vanishes. This would anyhow invite to pump-and-dump market manipulations \cite{Gatheral2010a}. \\
In section \ref{sec:aggr}, I will present a description of the trade formation which suggest that not the act of  buying or selling moves the price, but the underlying aggressiveness of the buyer and the seller. The aggressiveness' are result of the competition of the buyers (sellers) among their pairs. 

%%%%%%%%%%%%%%%%%%%%%%%%%%%%%%

\section{Slippage and accounting}
\label{sec:slippage}
In very general manner, one can model the market for a given asset by agents trading with another. These agents receive from their clients an exogenous parent orders to buy or sell a certain quantities of the asset. The purchase or selling of the asset will be described as a liquidation problem (a buyer have to liquidate a negative position, a seller liquidates a positive position). In practice, an agent can represent for instance a broker that deals with other broker over the phone for an OTC asset, or a execution algorithm that executes the given parent order in an electronic orderbook market. \\
The following framework is thus universal in the meaning that it can be applied to almost any kind of market. We do however not take into account that the client of the agent may change her mind during the execution and change (or cancel) the quantity to be executed; neither we consider market markers or high frequency trading where the parent orders do not exits.  These features might be easily taking into account, but do not present a fundamental interest for the present study. \\

\subsection{Agents, Wealth balance and Slippage}
We describe every participant of the market, who has the order to buy or sell a given quantity of the asset as an agent.  An agent $a$ appears at some time between the $i_a$th and the $(i_a+1)$th trade with a quantity $x_{i_a}^{a}$ to liquidate. For a buyer $x_{i_a}^{a} < 0$ while for a seller $x_{i_a}^{a}>0$. The initial reference price for agent $a$ is $ \tilde P^a :=P_{i_a}$, the price of the last trade. The mark-to-market valuation of the position to liquidate is thus initially $x_{i_a}^{a} \tilde P^a$. \\
In order to keep track of the trading of agent, we endow the agent with a cash account 
\footnote{This cash account can be considered as virtual in the meaning that it only a technical trick to evaluate the slippage. Nevertheless, the interpretation is that a client cedes a quantity of the asset to her agent in exchange of its mark-to-market value in cash. The agent liquidates then her position.}
with an initial value of  $C^{a}_{i_a} = -x_{i_a}^{a} \tilde P^a$.  In the framework of mark-to-market valuation, the initial total wealth (asset + cash) is then zero for all agents. Generally, after the $i$th trade, the wealth of the agent $a$ is given by
\begin{equation}
W_i^{a} := x^{a}_i P_i + C^{a}_i \quad \mathrm{for}\; i \ge i_a
\end{equation}
The asset position and the cash account change only if the agent participate in a trade. Let agent $s$ sell the quantity $q_i$ to agent $b$ at a price $P_i$ in the $i$th trade. The changes of the inventory of these agents are 
\begin{eqnarray*}
C^{b}_{i+1} - C^{b}_{i} &=& - q_i P_i \\
x^{p}_{i+1} - x^{b}_{i} &=&  +q_i \\
C^{s}_{i+1} - C^{s}_{i} &=&  +q_i P_i \\
x^{s}_{i+1} - x^{s}_{i} &=&  - q_i 
\end{eqnarray*}
Since a trade sets a new reference price, the wealth of all agents with $x^a_{i-1} \neq 0$ change by
\begin{equation}
\label{eqn:deltaW}
\Delta W_i^{a} := W_i^{a} - W_{i-1}^{a} = (x^{a}_i P_i - x^{a}_{i-1} P_{i-1})  - (x_i - x_{i-1}) P_i
\end{equation}
The first therm corresponds to holding the asset and the second term is the change of the cash account.  \\

Let us suppose that the agent has liquidated her position at finite $k_a := \inf \{i > i_a | x^{a}_i  = 0\}$. Setting $\Delta x^a_i := x^a_i - x^a_{i-1}$, the final wealth writes than as 
\begin{equation}
\label{eqn:sum}
W_{k_a}^{a} =  \sum_{i=i_a+1}^{k_a} \Delta W^a_i = -  \sum_{i=i_a+1}^{k_a} \Delta x^a_i (P_i - P^a)  = - TR^{a}
\end{equation}
We can therefore identity the final wealth of the agent with the transaction cost . For $k > k_a$, the wealth do not further evolve because $\Delta x^a_k = 0$.  \\
One might call $ - \Delta W^a_i$ the trade-by-trade-slippage of agent $a$. The total slippage is simply given by the sum over the trade-by-trade-slippages. \\

Notice furthermore the wealth change of the agent (equation \ref{eqn:deltaW}) can be written as the performance of the remaining position (position times return): 
\begin{equation}
\Delta W_i^{a} = x^{a}_{i-1} (P_i-P_{i-1})
\end{equation}
The slippage, usually seen as the difference of the achieved trade prices and the initial reference price, can also be interpreted as the gain (or rather loss) of the position until complete liquidation. Slippage is the { \it cost of waiting }. Note that this do not imply that trading as fast as possible minimizes slippage because the own trading has an impact on the price changes.  \\

%%%%%%%%%%%%%%%%%%%%%%%%%%%%%%

\subsection{Market average}
Let us now turn towards the whole market. Let $\mathcal{A}_i$ be the set of all agents present on the market after the $i$th trade, and  $\mathcal{S}_i = \{ a \in \mathcal{A}_i | x_i^{a} > 0 \}$ the set of sellers and $\mathcal{B}_i = \{ a \in \mathcal{A}_i | x_i^{a} < 0 \}$ the set of buyers. The total outstanding quantities to sell $V^{\mathcal{S}}$ and to buy $V^{\mathcal{B}}$ are
\begin{eqnarray*}
V^{\mathcal{S}}_i =& \sum_{s \in \mathcal{S}_{i}} x^{s}_{i} &>0\\
V^{\mathcal{B}}_i =& \sum_{ \in \mathcal{B}_{i}} |x^{b}_{i}| &>0
\end{eqnarray*}
Calling $N_i$ the number of active agents after the $i$th trade, the average change in wealth is the simply given by
\begin{equation}
\label{eqn:marketavg}
\frac{1}{N_i}\sum_{a \in \mathcal{A}} \Delta W^a_i
=  - \frac{1}{N_i} ( V^{\mathcal{B}}_{i-1} - V^{\mathcal{S}}_{i-1} ) (P_i-P_{i-1})
\end{equation}
The cumulated change of the wealth of all agents, the trade-by-trade slippage, it thus given by the product of the price change and the imbalance of the outstanding buy and sell volumes. One would expect that at least under normal circumstances, the temporal averages of the volume imbalance as well as the price changes are close to zero. But the slippage is given by their correlation. And this correlation is expected to be different from zero since an imbalance in buy and sell pressure is expected to impacts the price as I will illustrate in the following.\\
Hence, we can understand the slippage not only as the cost of waiting. It is due to {\it fluctuations in supply and demand } and the resulting price changes. \\

The trade-by-trade slippage is however not the slippage for a given agent. The agent will accumulate her trade-by-trade slippage until the liquidation of her position. As I will show in the following section, the cost of the correlation between price changes and volume imbalances is amplified by varying execution times. To do so, I will have to consider the competition between agents in more detail, which I will do in the following section.

%%%%%%%%%%%%%%%%%%%%%%%%%%%%%%
%%%%%%%%%%%%%%%%%%%%%%%%%%%%%%
%%%%%%%%%%%%%%%%%%%%%%%%%%%%%%

\section{Trade generation}
\label{sec:aggr}
We have seen that the average slippage corresponds to a correlation between the imbalance of outstanding buy and sell volumes and the price increase. While the price increase is directly observable, the outstanding volumes of all agents is unknown. Common sense implies that the outstanding volumes, the supply and demand, will influence the future price movement. \\
I will now present a generic description for the price formation where the price of the next trade is the result of a negotiation process. I will not consider a specific negotiation model (such as informal over the phone negotiations for OTC contracts, continuous auctions with orderbooks, \ldots), but consider a simple stochastic description as an example. 
% Conclusion ? 

\subsection{General formalism}
A trade is realized when a seller and a buyer agree on a price and a quantity. Let $P_{i}$ the price of the last trade that occurred at time $t_i$. This price is the reference common to all agents.  A seller agent $s$, at time $t > t_i$ would accept to sell one asset unity at a price $P_i - \epsilon^{s}(t)$ or higher, a buyer $b$ is willing to buy at a price lower or equal to $P_i+\epsilon^{b}(t)$. We call $\epsilon^{s}$ and $\epsilon^{b}$ the aggressiveness of the seller and the buyer respectively. Note that the aggressiveness may include very different things such as the need to buy or to sell or  personal believes about future price evolution \\
The agents trade if the price limit of the seller $s$ is lower or equal to the price limit of the buyer $b$, e.g. $\epsilon^{s} + \epsilon^{b} \ge 0$. Possible trade prices $P_{i+1}$ are then $P_i - \epsilon^{s}(t) \le P_{i+1} \le P_i + \epsilon^{b}(t)$. \\
In general, however, there are more than one buyer and one seller on the market. Naturally, it is the  buyer with highest bid and the seller with the lowest bid that trade with another. We shall therefore define the maximal aggressiveness' by 
\begin{eqnarray}
\epsilon_m^{\mathcal{B}}(t) &=& \max \{\epsilon^{b}(t) | b \in \mathcal{B} \} \\
\epsilon_m^{\mathcal{S}}(t) &=& \max \{\epsilon^{s}(t) | s \in \mathcal{S} \}
\end{eqnarray}
The condition for the occurrence of a trade at time $t^{\star}$ is then 
\begin{equation}
\epsilon_m^{\mathcal{B}}(t^{\star}) + \epsilon_m^{\mathcal{S}}(t^{\star}) \ge 0
\end{equation}
and the price of the trade is bounded by 
\begin{equation}
P_i -  \epsilon_m^{\mathcal{S}}(t^{\star}) \le P_{i+1} \le P_i + \epsilon_m^{\mathcal{B}}(t^{\star}) 
\end{equation}
Note that the competition of sellers (or buyers) translates into the rule that the only most aggressive agents participate in the trade. This description might apply to continuous prices as wall as to discrete prices (tick size). 

\subsection{Uniform unity trading agents}
\label{subsection:iid}
Let us consider the following simple model. Each agent has only one unity of asset to liquidate so that the outstanding volumes correspond the number of buyers and sellers. After a trade, one seller and one buyer disappear. They are replaced by two new agents so that the total number of agents $N$ remains constant. Let $N^{\mathcal{B}}$ be the number of buyers, $N^{\mathcal{S}} = N - N^{\mathcal{B}}$ the number of sellers. 
An arriving agent shall be a buyer with the probability
\begin{equation}
p^{\mathcal{B}} = \frac{N^{\mathcal{S}}}{N^{\mathcal{S}}+N^{\mathcal{B}}} 
\end{equation}
With this choice, there will be always at least one seller and one buyer.  The number of buyers and the number of sellers fluctuate around  $N/2$. \\

For the aggressiveness of the agents shall be given by following random procedure: An agent is chosen randomly, and her aggressiveness is set to a new random value given by a normal distribution with mean $\mu$ and standard deviation $\sigma$. If $\epsilon_m^{\mathcal{B}} + \epsilon_m^{\mathcal{S}} \ge 0$, then the most aggressive buyer trades with the most aggressive seller at a price $P_{i+1} = P_i + (\epsilon_m^{\mathcal{B}}(t) - \epsilon_m^{\mathcal{S}}(t)) / 2$. \\

\begin{figure}
\begin{center}
\includegraphics[width=8cm]{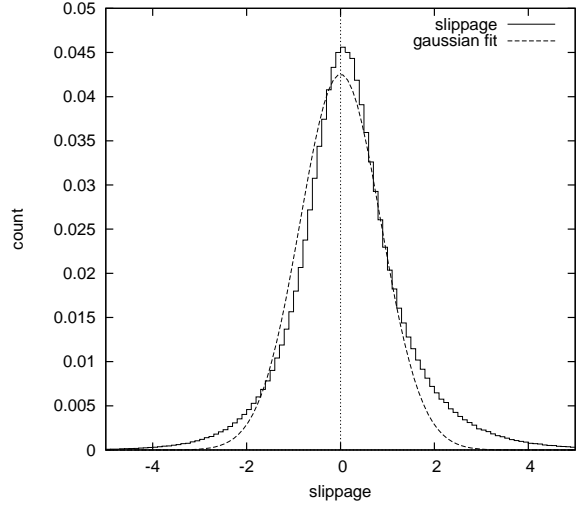}
\caption{The distribution of execution times. The average value is indicated by the vertical bar.}
\label{fig:gauss}
\end{center}
\end{figure}

\begin{figure}
\begin{center}
\includegraphics[width=8cm]{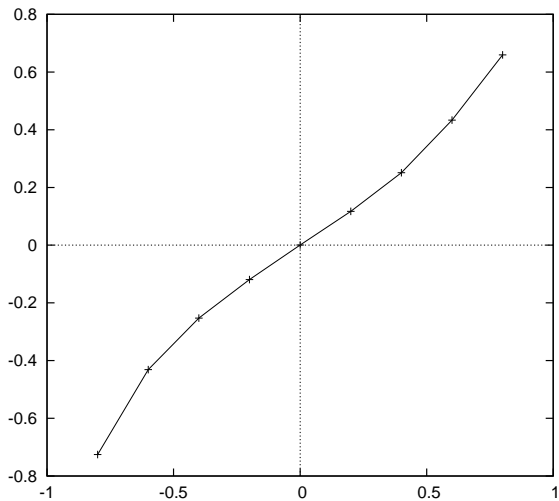}
\caption{The average price increase as a function of the buy-sell imbalance. }
\label{fig:impact}
\end{center}
\end{figure}

The figures show the results of a numerical simulation of this model with $N=10$, $\mu = -0.5$ and $\sigma=0.2$. Simulations with different values, or with random varying quantities to execute or different agent creation models give similar results. \\
Figure \ref{fig:gauss} shows the histogram of the slippages of the agents together with a Gaussian fit with zero mean as a guide. One observes that the average slippage is positive, and the distribution has a positive skewness. These properties can be understood easly with the remain figures. \\
Figure \ref{fig:impact} shows the average price increase between two trades conditioned on the imbalance of the outstanding volumes defined by
\begin{equation}
\lambda  = 2 \frac{ V^{\mathcal{B}} - V^{\mathcal{S}} }{V^{\mathcal{B}} + V^{\mathcal{S}}} -1
\end{equation} 
One clearly realizes the positive correlation between the price change and the volume imbalance. In this simple model, this correlation is explained by the distributions of the maximal agressiveness'. Let us for instance consider the cumulative distribution function for $\epsilon_m^{\mathcal{B}}$  : 
\begin{equation}
P(\epsilon_m^{\mathcal{B}} < \Theta ) = \prod_{b \in \mathcal{B}}  P(\epsilon^{b} < \Theta) =  P(\epsilon^{b} < \Theta) ^{N^{\mathcal{B}}}
\end{equation}
The maximal aggressiveness of the buyers increase thus with the number of buyers, and the maximal aggressiveness of the sellers with the number of sellers. More buyers than sellers therefore implies that the buy side aggressiveness is higher than the sell side aggressiveness and the expected price change is positive. \\

\begin{figure}
\begin{center}
\includegraphics[width=8cm]{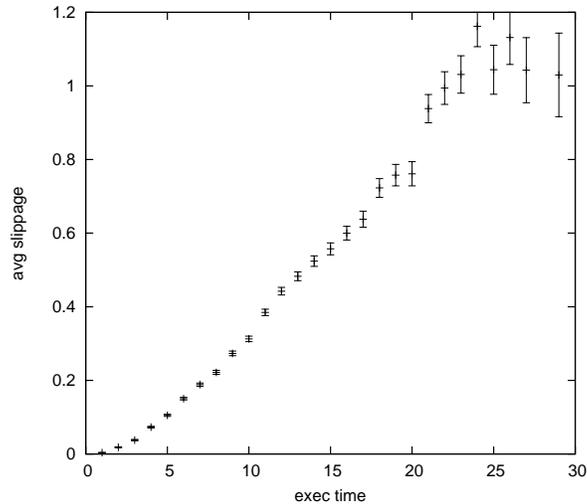}
\caption{The average slippage as a function of the execution time.}
\label{fig:avg}
\end{center}
\end{figure}

The skewness of the slippage distribution can be explained by the execution times. Figure \ref{fig:avg} shows the average slippage conditioned on the execution time $\tau$ (the number of trades occurred after the appearance of the agent until the liquidation of her position). For all $\tau$, the average slippage is positive. It increases clearly with the execution time. The reason for this is the following: Let us suppose that there are more sellers than buyers. Due to the negative price pressure, a buyer will the profit from price decrease but will execute her order quickly because of the weak competition among the buyers. Note that the probability $p_{trade}^{b}$ that buyer $b$ participates in the next trade is simply given by 
\begin{equation}
p_{trade}^{b} = \frac{1}{N^{\mathcal{B}}}
\end{equation}
In the extreme case where there is only one buyer, it it clear that she will participate in the next trade. On the other hand, a seller will suffer from the price decrease, and furthermore, because of the larger number of buyers, it will take her a longer time to liquidate her position. There is thus in in general a small gain for the minority and a larger loss for the majority. \\

\section{Conclusion}
Trading costs have a uncontested importance in finance, and in particular over the last years, hedge funds and banks have invested enormous sums into infrastructure and research such to control and optimize their executions. Electronic platforms and dark pools are increasing used in the hope to minimize slippage. The arguments presented in this papers however suggest that positive slippage is intrinsic to trading seen as a negotiation process between agents. It is beneficial to be in the minority, but unfortunately, you will find yourself more often in the majority and thus in a stronger competition. \\
I argued that slippage is the direct consequence of the correlation between the supply and demand imbalance and the price change. And since this correlation is nothing but the price formation for an asset, one could ask where a zero slippage market would lead. \\
The model for the aggressiveness of the agents used for illustration in section \ref{subsection:iid} is of course to simple to be realistic, but it has the advantage that everything  about the agents is known. Since one can not expect to obtain data containing the intentions and strategies of al agents even for a short period of time, it seems promising to study more complex (realistic) models of trading agents theoretically to gain deeper inside in the mechanics of the financial markets. 

\bibliographystyle{plain}
\bibliography{bibleo.bib}

\end{document}